\documentclass{WileyMSP-template}

\usepackage{caption}
\usepackage{amsmath,amssymb}
\usepackage{float}
\usepackage{geometry}
\usepackage[dvipsnames]{xcolor}

\begin{document}
\pagestyle{fancy}
\rhead{\includegraphics[width=2.5cm]{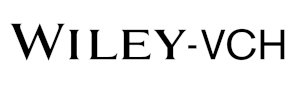}}

\title{Direct Measurement of the Singlet Lifetime and Photoexcitation Behavior of the Boron Vacancy Center in Hexagonal Boron Nitride}
\maketitle

\renewcommand{\thefootnote}{\fnsymbol{footnote}}
\author{Richard A. Escalante$^{1,2}$\footnotemark[1]}
\author{Andrew J. Beling$^{1,3}$\footnotemark[1]}
\author{Daniel G. Ang$^{1}$}
\author{Niko R. Reed$^{1,}$$^{3,}$$^{4}$}
\author{Justin J. Welter$^{1,}$$^{2}$}
\author{John W. Blanchard$^{1}$}
\author{Cecilia Campos$^{5}$}
\author{Edwin Coronel$^{5}$}
\author{Klaus Krambrock$^{6}$}
\author{Alexandre S. Leal$^{7}$}
\author{Paras N. Prasad$^{5}$}
\author{Ronald L. Walsworth$^{1,}$$^{2,}$$^{3,}$$^{4}$}

\begin{affiliations}

$^{1}$Quantum Technology Center, University of Maryland, College Park, MD 20742, USA\\
$^{2}$Department of Electrical and Computer Engineering, University of Maryland, College Park, MD 20742, USA\\
$^{3}$Department of Physics, University of Maryland, College Park, MD 20742, USA\\
$^{4}$Joint Quantum Institute, University of Maryland, College Park, MD 20742, USA\\
$^{5}$Institute for Lasers, Photonics and Biophotonics, and the Department of Chemistry,\\
University at Buffalo, State University of New York, Buffalo, NY 14260, USA\\
$^{6}$Department of Physics, Federal University of Minas Gerais (UFMG), Belo Horizonte, MG, Brazil\\
$^{7}$Nuclear Technology Development Center (CDTN), Belo Horizonte, MG, Brazil

Email Address: \texttt{rescala1@umd.edu} (R.A.E.), \texttt{abeling@umd.edu} (A.J.B.), \texttt{walsworth@umd.edu} (R.L.W.)
\end{affiliations}

\footnotetext[1]{These authors contributed equally to this work.}

\begin{abstract}
Optically active spin defects in van der Waals (vdW) materials are a promising platform for quantum sensing, potentially enabling shorter standoff distances than defects in diamond and thus improved measurement signal-to-noise ratio (SNR) and spatial resolution. The most studied such defect is the negatively charged boron vacancy center ($V^{-}_{B}$) in hexagonal boron nitride (hBN), yet many of its electronic and spin transition rates and branching ratios remain unknown. Here, we use time-resolved photoluminescence (PL) measurements with a nanosecond rise-time 515 nm laser to directly measure the singlet state lifetime of a $V^{-}_{B}$ ensemble in neutron-irradiated, sub-micron flakes of hBN. We perform this measurement on 16 flakes at room temperature and obtain an average lifetime of 15(3) ns. Additionally, we probe the PL dynamics of thermal and optically polarized electronic spin distributions of the $V^{-}_{B}$ ensemble in a sub-micron hBN flake, and fit our results to a 9-level model to extract electronic transition rates. Lastly, we present PL measurements that potentially indicate optically-induced conversion of $V^{-}_{B}$ to another electronic state, or possibly the neutral charge state ($V^{0}_{B}$), in neutron-irradiated hBN flakes of size $>$ 1 $\mu$m.

\end{abstract}

\section{Introduction}

Research on optically-active spin defects in solid-state materials has grown significantly in recent decades, enabling advances in quantum technologies, including single quantum emitters \cite{Aharonovich2016}, quantum-based memories \cite{Bradley2022,Sukachev2017,Parthasarathy2023}, and quantum sensing \cite{Maze2008,Hsieh2019,Barry2020,Wang2023}. The nitrogen vacancy (NV) center in diamond has emerged as the leading solid-state platform in quantum sensing, particularly for magnetometry\cite{Rondin2014,Barry2020,Sekiguchi2024}, as well as for electrometry\cite{Bian2021}, thermometry\cite{Wang2015}, strain sensing \cite{marshall_high-precision_2022}, and inertia measurement\cite{zhaoInertialMeasurementSolidstate2022}. However, NVs and other solid-state defects are typically constrained by the need to reside at least several nanometers from the host-material surface, as they become unstable near the material boundary \cite{Kumar2024}. A defect that remains stable at or near the surface of a two-dimensional (2D) material could substantially reduce the stand-off distance to a given sensing target, potentially providing great value in diverse applications including: nuclear magnetic resonance (NMR) of nanoscale samples \cite{daly2025prospectsultralowmassnuclearmagnetic, Vaidya2023}; local pressure imaging in diamond anvil cells\cite{he2025}; and electronic circuit performance evaluation and authenticity validation.\cite{ASHOK2022}

\begin{figure*}[h]
  \centering
  \includegraphics[width=0.9 \textwidth]{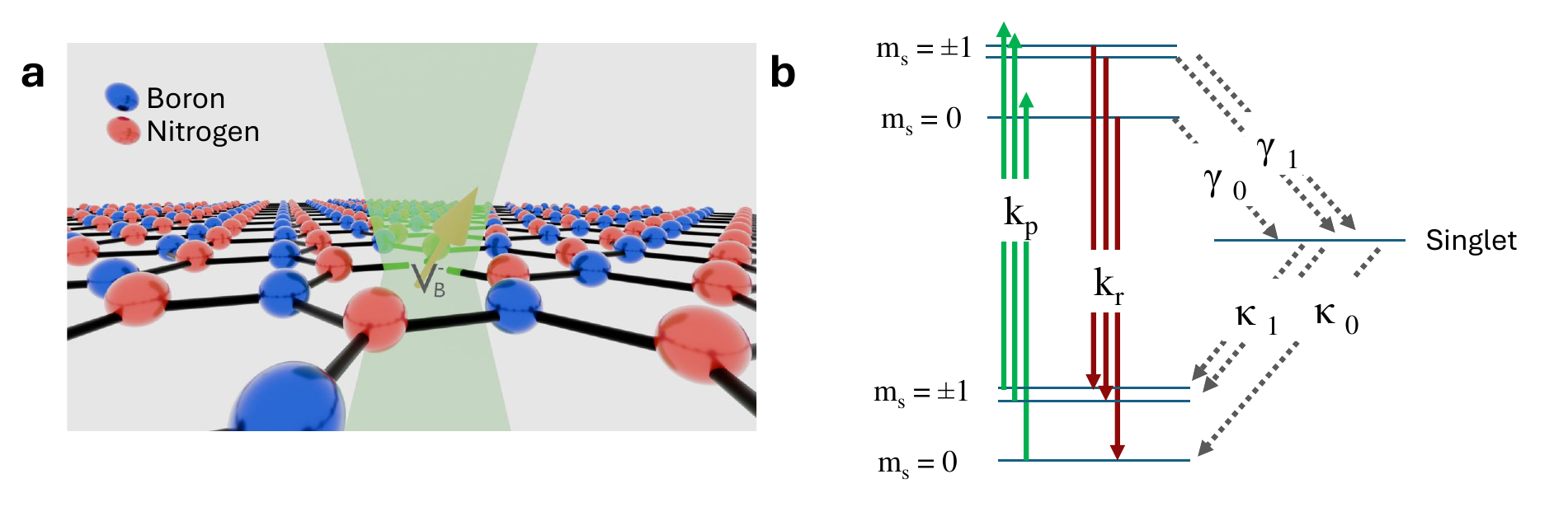}
  \caption{a) Illustration of $V^{-}_{B}$ defect in hBN. b) $V^{-}_{B}$ energy level diagram showing the 7-level model.}
  \label{Fig_1}
\end{figure*}

Hexagonal boron nitride (hBN) is a van der Waals (vdW) material and promising spin defect host due to its wide band gap, chemical and thermal stability, and optical properties at room temperature. Out of many optically active spin defects within the 2D hBN lattice\cite{Glushkov2022}, which can be created by both irradiation and defect center implantation, the negatively charged boron vacancy ($V^{-}_{B}$) (Fig. \ref{Fig_1}a) has attracted interest for sensing due to its similarity to the NV, including a ground state electronic spin triplet with S=1, optically detected magnetic resonance (ODMR), and optically induced spin polarization, as well as its good optical and spin properties in 3-5 layer flakes of hBN.\cite{Mathur2022,Guo2022,Durand2023} $V^{-}_{B}$ ensembles have been used for sensing of magnetic fields, pressure, and temperature, with electronic spin coherence times $\sim$0.1-1 $\mu$s\cite{liu2021,healey2023,mu2025,he2025,Gottscholl2021,gao_high-contrast_2021}. Additionally, $V^{-}_{B}$ can be easily created in hBN using a variety of irradiation techniques.  \cite{Kianinia2020,Gao2021,Murzakhanov2021,Gottscholl2020,Toledo2018}

The $V^{-}_{B}$ energy level structure is typically approximated by a 7-level model, including ground and excited state spin triplets and a metastable spin singlet state (Fig. \ref{Fig_1}b).\cite{Provost2024,lee2024intrinsichighfidelityspinpolarization} However, despite extensive studies, $V^{-}_{B}$ photoexcitation dynamics and properties are not fully understood. In particular, the lifetime of the metastable singlet state ($\tau_s $) has never been directly measured. (In the 7-level model of Fig. \ref{Fig_1}b, the singlet lifetime is given by the inverse of the intersystem crossing rates: $\tau_s = (2\kappa_1 + \kappa_0)^{-1}$.) Previous indirect measurements and theoretical studies report conflicting values (Table~\ref{tab:lifetimes}).\cite{Reimers2020,Baber2021, Whitefield2023, Provost2024, lee2024intrinsichighfidelityspinpolarization} For example, Clua-Provost et al. experimentally determined $\tau_s$ = 18(3) ns\cite{Provost2024}, while Whitefield et al. obtained $\tau_s$ = 30(1) ns\cite{Whitefield2023}, and Reimers et al. predicted $\tau_s$ = 2 s based on \textit{ab initio} Density Functional Theory (DFT) calculations.\cite{Reimers2020} These and other previous experimental studies indirectly extract $\tau_s$ from a fit of measured photoluminescence (PL) to an energy level model, switching the incident laser light using standard acousto-optic modulators (AOMs) that have a rise-time significantly longer than the expected $\tau_s \sim$ 10-30 ns.\cite{Provost2024,lee2024intrinsichighfidelityspinpolarization,Reimers2020,Baber2021,Whitefield2023} A direct measurement of $\tau_s$ requires the use of the pulse recovery protocol, with switching of the incident laser light on a timescale fast compared to $\tau_s$, as was previously used to measure the singlet lifetime of the NV center.\cite{Manson2006,Robledo2011}. Here, we perform a direct measurement of $\tau_s$ with the pulse recovery method, utilizing a digitally modulated laser with a fast rise time ($\lesssim 2.5$ ns, Hubner Cobolt 06-MLD). We perform the measurement on 16 neutron-irradiated, sub-micron-size hBN flakes at room temperature, obtaining a mean value of $\tau_s$ = 15(3) ns.

\begin{table}[h]
\centering
\begin{tabular}{|l|c|}
\hline
\textbf{Reference} & \textbf{$\tau_s$ (ns)} \\ \hline
Reimers et al., DFT calculation [\citen{Reimers2020}] & 1.8$\times 10^9$ \\ \hline
Baber et al., indirect measurement [\citen{Baber2021}] & 12.7 \\ \hline
Whitefield et al., indirect measurement [\citen{Whitefield2023}] & 30(1) \\ \hline
Clua-Provost et al., indirect measurement [\citen{Provost2024}] & 18(3) \\ \hline
Lee et al., indirect measurement [\citen{lee2024intrinsichighfidelityspinpolarization}] & 31.5 \\ \hline
This work, direct measurement & 15(3) \\ \hline
This work, 7-level model & 14.4(4)\\ \hline
This work, 9-level model & 17.2(6)\\ \hline
\end{tabular}
\caption{$V^{-}_{B}$ singlet lifetime as reported in prior studies and directly measured in this work. Uncertainties in $\tau_s$ measurements are not reported in [\citen{lee2024intrinsichighfidelityspinpolarization}] and [\citen{Baber2021}].}
\label{tab:lifetimes}
\end{table}

$V^{-}_{B}$ electronic and spin transition rates are also of interest for understanding the spin defect's quantum dynamics; see Fig. \ref{Fig_1}b. Under optical excitation, the transition rate from the ground-state triplet to the excited-state triplet is spin-conserving and given by $k_p$, which is linearly proportional to the laser power $P$ such that $k_p = P k_{p_0}$ \cite{Whitefield2023, Provost2024}. The excited state triplet can undergo a spin-conserving radiative decay back to the ground state at rate $k_r$. It should be noted that excited state lifetime measurements include contributions from both radiative and non-radiative rates, where the non-radiative rates are spin-dependent and denoted by $\gamma_0$ and $\gamma_1$. Additionally, the first excited state transition is optically forbidden for unstrained hBN, and thus only occurs in the presence of phonons or hBN lattice static distortions (e.g., local strain and other defects); consequently, the quantum efficiency of the $V_B^-$ defect is very low.\cite{LIBBI2022, ABDI2018} \textit{Ab initio} calculations using DFT predict $k_r$ $\approx$ $\frac{1}{11}$ MHz\cite{Reimers2020} and $k_r$ $\approx$ $\frac{1}{20}$ MHz,\cite{ivady_ab_2020} which is about 3 orders of magnitude weaker than the NV- radiative decay rate \cite{AudeCraik}. Recent measurements using F{\"o}rster resonance energy transfer (FRET) report $k_r = 0.135$ MHz $\pm$ $0.068$ MHz \cite{jules2025nonradiativeenergytransferboron}, in reasonable agreement with DFT predictions. In [\citen{CluaProvost2024}], $k_r$ is inferred from excited-state lifetime measurements to be equivalent for all 3 spin-states of $V^{-}_{B}$. Like the NV center, the $V^{-}_{B}$ center undergoes a spin-dependent non-radiative decay from the excited triplet to a metastable singlet. The excited $m_s$ = $\pm$1 levels are more likely than the excited $m_s$ = 0 level to decay through this pathway, leading to spin-dependent PL brightness. Population in the singlet state decays preferentially to the $m_s$ = 0 ground state, causing spin polarization to this brighter PL state, similar to the NV. (See [\citen{Barry2020}] for a thorough discussion of NV spin-dependent PL brightness.)

To determine $V^{-}_{B}$ transition rates between energy levels, we perform optical pump-probe measurements of thermal and polarized spin states in a single sub-micron hBN flake, and fit time-resolved PL data to the 7-level model, allowing us to extract the transition rates, $\kappa_0$, $\kappa_1$, and $k_{p0}$, as well as the ratio $\frac{\gamma_0}{\gamma_1}$. Nanosecond rise-time laser pulses allow for both direct measurement of the singlet lifetime $\tau_s$ and also finer resolution of the sharp PL peak at the beginning of a pulse. The impact of laser rise-time on time-resolved PL is summarized in the Supporting Information (SI). We also study flake- and laser-power-dependent changes to the $V^{-}_{B}$ emission spectrum and PL quenching recovery on time scales longer than the measured singlet lifetime. We find that the time-resolved PL data are better represented by a 9-level model of the $V^{-}_{B}$ energy structure, with additional laser-power dependent electronic transition rates between the original 7 states and the 2 additional states. Lastly, we report distinct $V^{-}_{B}$ photophysical behavior in large hBN flakes ($>$1 $\mu$m), including a long PL decay ($\sim$1 $\mu$s) for a laser pulse pair with a relatively long wait-time between pulse pair repetition (1 $\mu$s), with this long PL decay not observed for laser pulse pairs with a relatively short wait-time (100 ns). Additionally, continuous laser excitation and higher laser powers produce an increase in PL intensity for large hBN flakes within the $\sim$675-775 nm range. This behavior may arise from optically-induced recombination dynamics between defect charge states $V^{-}_{B}$ and $V_B^0$.

\section{Results and discussion}

\begin{figure*}[h]
  \centering
  \vspace{0.5cm}
  \includegraphics[width=0.9 \textwidth]{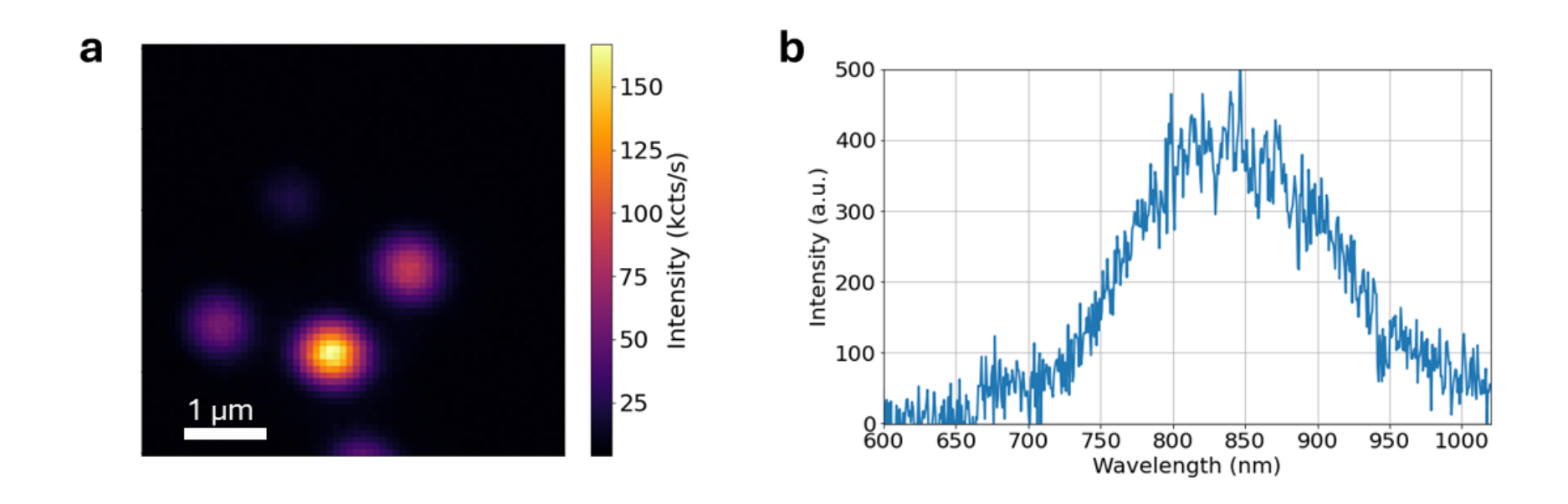}
  \caption{a) Example scanned-confocal PL map of neutron irradiated flakes of hBN. b) Typical $V^{-}_{B}$ PL emission spectrum for a sub-micron hBN flake with peak emission centered near 825 nm.}
  \label{Fig_2a}
\end{figure*}

We investigate the PL behavior of ensembles of $V^{-}_{B}$ defects in irradiated hBN flakes at room temperature with a homemade confocal microscope (details in Experimental Methods). Figure \ref{Fig_2a}a shows an example diffraction-limited PL scan of a typical set of flakes. Using a nanosecond rise-time laser with excitation power of approximately 13.6 mW at 515 nm, we identify flakes with a PL emission spectrum like that shown in Fig. \ref{Fig_2a}b, with peak emission centered near 825 nm.\cite{Gottscholl2020} We report results on smaller hBN flakes ($<$1 $\mu$m) separate from those on larger flakes ($>$1 $\mu$m) because larger flakes exhibit a laser-power dependent change in the PL emission spectrum, possibly indicating photoconversion of $V^{-}_{B}$ to other charge states such as $V^{0}_{B}$, which can interfere with the measurement of $\tau_s$ (see SI Fig. S2).

\subsection{Singlet lifetime measurement}
\label{sec:singletlifetimemeas}
\begin{figure*}[h]
  \centering
  \includegraphics[width=0.9 \textwidth]{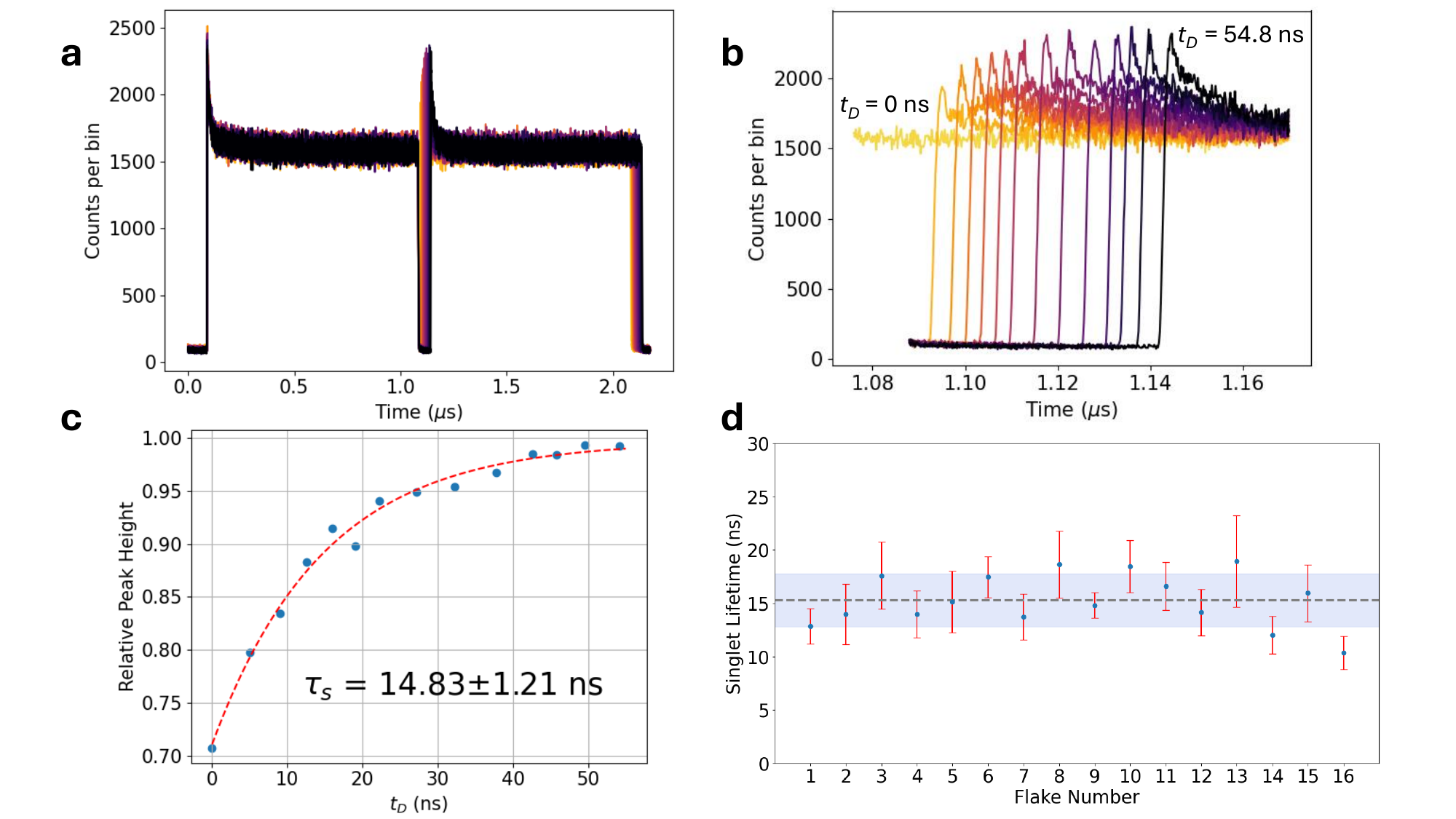}
  \caption{a) Example signal-averaged PL time traces from a sub-micron hBN flake induced by pairs of 1 $\mu$s laser pulses, with spacings between pulses (dark time $t_D$) ranging from 0 ns to 54.8 ns. b) Zoomed-in display of PL time-trace data in a) from the second laser pulse showing PL peak height recovery for different $t_D$. c) Measured relative PL peak height as a function of $t_D$ for one sub-micron hBN flake. Red dashed curve shows exponential fit yielding metastable singlet state lifetime $\tau_s$ for this flake. d) Fitted $\tau_s$ values and uncertainties from PL time-trace measurements of 16 sub-micron hBN flakes. Gray dashed line designates mean lifetime from all the flakes and shaded region is standard deviation: $\tau_s$ = 15(3) ns.}
  \label{Fig_2}
\end{figure*}

To perform the $V^{-}_{B}$ singlet lifetime measurement for an individual sub-micron hBN flake, we apply pairs of closely spaced 515 nm laser excitation pulses and measure the resulting PL time trace (Fig. \ref{Fig_2}a). The pulses are 1 $\mu$s long and the spacing between pulses (i.e., dark time $t_D$) is varied from 0 to 54.8 ns;  measurements are enabled by use of the nanosecond rise-time laser. The wait-time between pulse pairs is 100 ns, which we find to be sufficient to allow near-complete population decay to the electronic ground state. This wait-time is chosen based on the experimentally determined singlet lifetime of approximately 18 ns \cite{Provost2024}, ensuring near-total ground-state population recovery within 100 ns; and is also significantly shorter than the previously measured $V_B^-$ spin-lattice relaxation time $T_1$ $\sim$13–19 $\mu$s \cite{Durand2023,Provost2024,Liu2022, Gottscholl2021-2} and possible recovery dynamics discussed in sections \ref{sec:transitionrates} and \ref{sec:largehbn}, ensuring minimal effect from spin-lattice relaxation and additional relaxation processes during the measurement. For each value of $t_D$, we repeat the measurement protocol for $\sim$5 minutes and signal average the results, with an entire set of measurements for all $t_D$ values requiring $\sim$1 hour. We observe a clear recovery in PL peak height with increasing $t_D$, due to excited state population decaying to the ground state before the second pulse (Fig. \ref{Fig_2}b). To extract the metastable singlet state lifetime $\tau_s$, we take the ratio of the (signal-averaged) height of the second PL peak to that of the first peak and plot it as a function of the dark time, $t_{D}$ (Fig. \ref{Fig_2}c). The measurement is then repeated on 16 hBN sub-micron flakes from the same batch (Fig. \ref{Fig_2}d). Averaging results from all flakes, we obtain a mean $\tau_s$ = 15(3) ns; this value is consistent with Clua-Provost et al.,\cite{Provost2024} which indirectly inferred $\tau_s$ = 18(3) ns. The contrast of observed PL peak height to the steady state PL value under laser excitation can give a lower bound estimate on $V^{-}_{B}$ polarization to the singlet state.\cite{Bayliss2020} We report this value for the 16 studied sub-micron hBN flakes in Fig. S8 of the SI.

The modest observed variability in $\tau_s$ values from the studied set of 16 sub-micron hBN flakes may be a result of differences in flake properties. A prior study \cite{CluaProvost2024} observed variations in excitation laser light absorption, PL collection efficiency, and Purcell factor depending on flake thickness and defect depth. Another possibility is variation in the susceptibility of different flakes to optically-induced changes in defect charge state; e.g., if the other charge state is dark (i.e., does not produce observable PL), it would act as an additional shelving state and affect determination of $\tau_s$ from PL recovery measurements.

\subsection{Measurements of transition rates}
\label{sec:transitionrates}
\begin{figure*}[h]
  \centering
  \includegraphics[width=0.9 \textwidth]{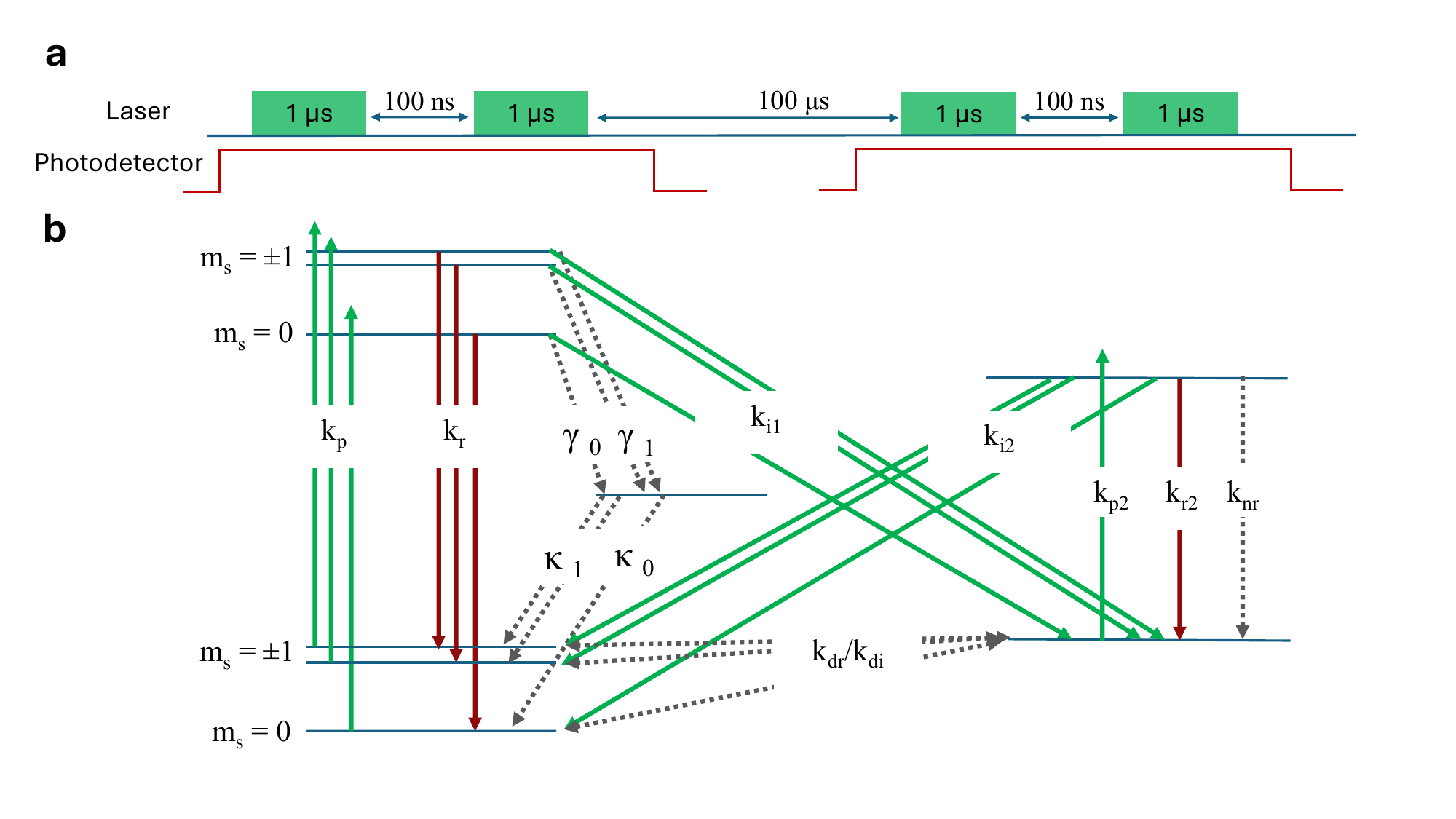}
  \vspace{-30pt}
  \caption{a) Measurement sequence used to probe PL from the thermal and polarized electronic spin distributions of ensembles of $V^{-}_{B}$ in a neutron irradiated flake of hBN. A pair of 1 $\mu$s laser pulses is applied with a dark time spacing of 100 ns. The 1 $\mu$s pulse duration allows partial spin polarization to the singlet state. The 100 ns dark time is sufficient for significant decay from the singlet state, preferentially to the $m_{s}$ = 0 ground state (see Fig. 1b), at which point the second laser pulse induces PL from the polarized spin distribution. The 100 $\mu$s  wait-time between pulse pairs is significantly longer than the $V^{-}_{B}$ spin $T_1$ to allow for full depolarization  to a thermal spin distribution. b) Modified $V^{-}_{B}$ energy level diagram (9-level model) including 2 additional states to better describe thermal and polarized spin distribution PL data. Additional conversion pathways between the 2 additional states and the ground and singlet states of the 7-level model do not measurably influence fits to data and are thus are omitted.}
  \label{Fig_4}
\end{figure*}

\begin{figure*}[h]
  \centering
  \includegraphics[width=0.9 \textwidth]{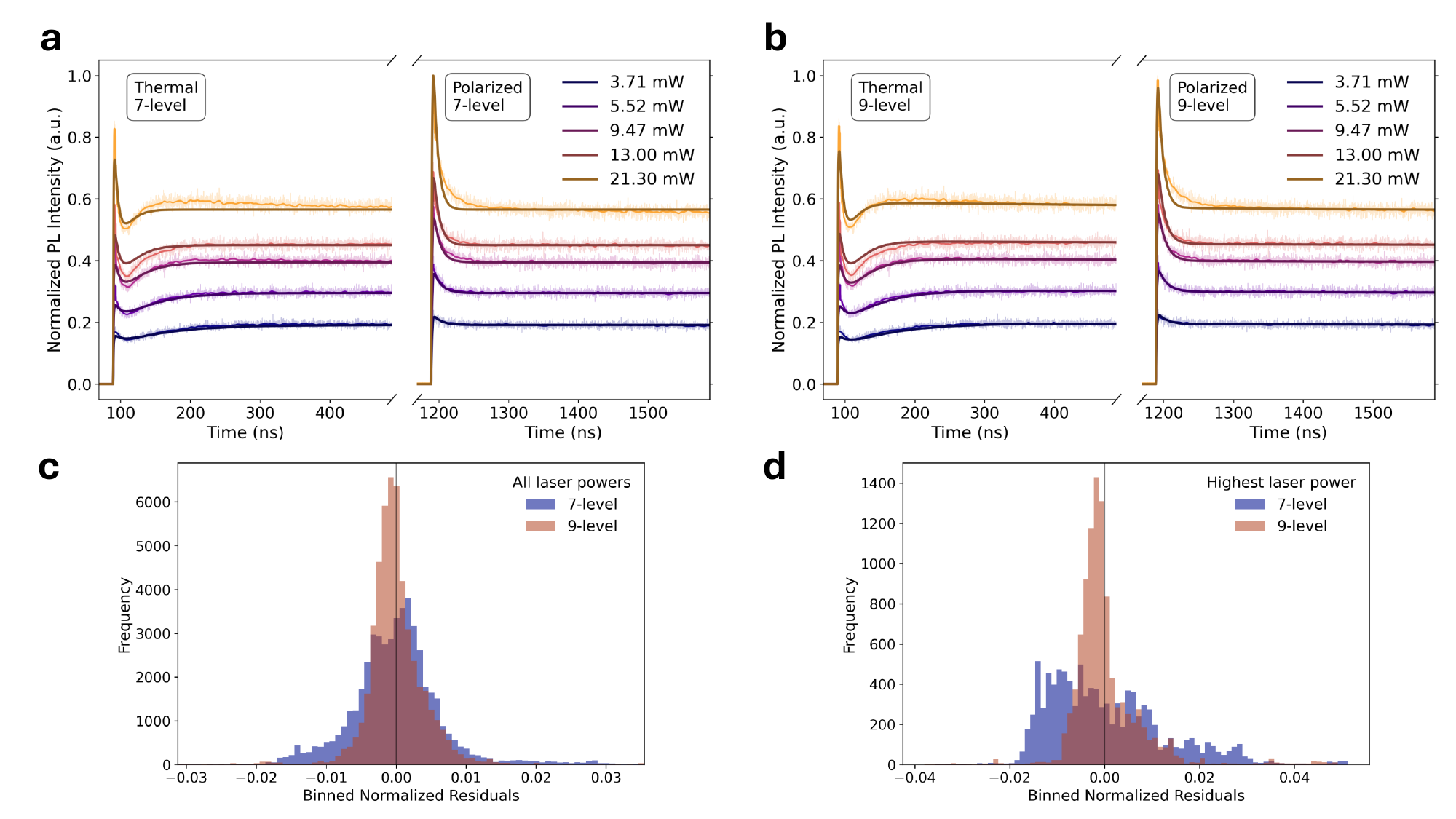}
  \vspace{-10pt}
  \caption{a), b) Measured PL time traces for a sub-micron hBN flake at room temperature, following the sequence in 4a), showing thermal (left) and polarized (right) $V_{B}^{-}$ spin distributions and fits to the a) 7-level model, b) 9-level model. For each laser power, PL measurements are repeated and signal averaged over $\sim$15 minutes, with an entire set of measurements for all laser powers requiring $\sim$ 1.5 hours. Raw data is smoothed with a mean shift filter and shown with a semi-opaque color scale compared to the smoothed data and fits (solid lines). c) Histograms of fit residuals for all five laser powers, comparing the 7-level and 9-level model fits to the data. The distribution of 9-level model residuals is more sharply peaked around 0, indicating a better fit. d) Histograms of fit residuals for the highest laser power. The distribution of 7-level model residuals has a wider spread compared to c); whereas the distribution of 9-level model residuals remains sharply peaked around 0.}
  \label{fig:Fig_5_Rev}
\end{figure*}

To determine the $V^{-}_{B}$ energy level transition rates, we measure time-resolved PL from a single sub-micron hBN flake as a function of laser power and initial ground-state spin distribution. We apply pairs of 1 $\mu$s laser pulses, with a dark time between pulses of 100 ns and a wait-time between pulse pairs of 100 $\mu$s. $V^{-}_{B}$ PL is measured during the application of each laser pulse (see measurement sequence in Fig. \ref{Fig_4}a). The wait-time is chosen to be long compared to the $V^{-}_{B}$ spin-lattice relaxation time $T_1$ $\sim$13--19 $\mu$s  \cite{Liu2022,Gottscholl2021-2, Provost2024, Durand2023}, thereby allowing the $V^{-}_{B}$ spin population to fully relax to a thermal distribution between each laser pulse pair. The 100 ns dark time between laser pulses is short compared to $T_1$, but long compared to the measured singlet lifetime $\tau_s$, thereby allowing study of the photophysics of a nearly spin-polarized $V^{-}_{B}$ ensemble. The measured photophysical response, i.e., PL dynamics, of the thermal and polarized $V^{-}_{B}$ electron spin distributions are shown in Fig. \ref{fig:Fig_5_Rev}a. We repeat the measurements for five laser powers between 3.71 and 21.3 mW (measured at the focus of the beam after the objective). Signal-averaged PL from the first $\sim$200 ns of each laser pulse is shown in Fig. \ref{fig:Fig_5_Rev}a, with full PL data given in the SI.  For both $V^{-}_{B}$ distributions, there is an initial PL peak followed by PL quenching and equilibration to a steady-state value, resulting from rapid optical initialization to the photoluminescent excited states and then partial population decay to the dark singlet state. For the thermal distribution, PL recovers from a minimum to a steady-state value as spin population is polarized to the relatively brighter $m_s=0$ state. Optical initialization to $m_s=0$ occurs faster for higher laser powers, with steady-state PL achieved within $\sim$100-300 ns. The exception to this behavior is at the highest power (21.3 mW), where a decay in counts occurs on longer time scales (SI Figure S5).

For both thermal and polarized spin distribution measurements, we first fit the time-resolved PL data to the 7-level model of the $V^{-}_{B}$ electronic and spin energy levels (see Fig. \ref{Fig_1}b). To bound the parameters more reliably and reduce sensitivity to local minima, we use a multi-start fitting routine based on Latin hypercube sampling (LHS) of the parameter space (see [\citen{mckayComparisonThreeMethods2000}] and [\citen{doi:10.1021/jacs.5c05505}]). We keep $k_{p0}$, $\kappa_0$, $\kappa_1$, and the ratio between non-radiative decay-rates $\frac{\gamma_0}{\gamma_1}$ as free parameters in the model and set the radiative decay rate $k_r$ to the DFT simulation value \cite{Reimers2020}. 
$k_{p_0}$ is a fit parameter that determines the scaling of the optical pumping rate $k_p$ with laser power, $P$, such that $k_p=P\cdot k_{p_0}$. The LHS fitting procedure indicates that the absolute magnitudes of $\gamma_0$ and $\gamma_1$ are not uniquely constrained by the PL data, whereas the ratio $\frac{\gamma_0}{\gamma_1}$ converges robustly; therefore, either $\gamma_0$ or $\gamma_1$ must be constrained by an independent measurement. Hence, we fix $\gamma_0$ determined from a literature average of directly measured $m_s=0$ excited state lifetimes ($\tau_0$) \cite{Provost2024, Baber2021, lee2024intrinsichighfidelityspinpolarization}. Using $\tau_0=\frac{1}{k_r+\gamma_0}$ and $k_r<<\gamma_0$ \cite{ivady_ab_2020, Reimers2020, Provost2024, jules2025nonradiativeenergytransferboron}, yields $\gamma_0=1/\tau_0$. Fit parameter uncertainties are calculated by adding in quadrature the statistical uncertainty of the fit, obtained from bootstrap resampling of the data, and the systematic uncertainty estimated from jackknife resampling\cite{wu_jackknife_1986_wildbootstrap} of data at different laser powers. Details of PL data fitting to the 7-level model are given in the SI, with resulting transition rates summarized in Table 2 and compared to past results.

The fit results demonstrate that the 7-level model accurately accounts for normalized PL measurements at low laser powers. 
From the reverse intersystem crossing rates $\kappa_0$ and $\kappa_1$, we extract the singlet lifetime $\tau_s=\frac{1}{\kappa_0+2\kappa_1}$ = 14.4(4) ns, which is consistent with our independent measurement of $\tau_s$ in section \ref{sec:singletlifetimemeas}. However, the 7-level model fails to give good fits to data at high laser powers. At the highest laser power (21.3 mW), the model does not account for the observed long-time PL quenching (decay) over the course of the first laser pulse (Fig.~\ref{fig:Fig_5_Rev} and SI Figure S5). 

\begin{table}[h]
\centering
\renewcommand{\arraystretch}{1.25}
\begin{tabular}{|l|c|c|c|c|c|c|c|}
\hline
\textbf{Reference} & \textbf{$k_{p_0}$ (MHz/mW)} & \textbf{$k_{r}$ (MHz)} & \textbf{$\gamma_0$ (MHz)} & \textbf{$\gamma_1$ (MHz)} & \textbf{$\kappa_0$ (MHz)} & \textbf{$\kappa_1$ (MHz)} \\ \hline
This work, 7-level model & 3.6(1) & 0.091 from [\citen{Reimers2020}]& $1.003$
 $\times 10^3$ (fixed) & 2.14(1) $\times 10^3$  & 58(1) & 5.8(2) \\ \hline
Baber et al. [\citen{Baber2021}] & Not stated & 0.091 from [\citen{Reimers2020}] & 1.01 $\times 10^3$ & 2.03 $\times 10^3$ & 20 & 29.4 \\ \hline
Whitefield et al. [\citen{Whitefield2023}] & Not stated & 8.8(3) $\times 10^2$ & 1.3(3) $\times 10^2$ & 1.15(3) $\times 10^3$ & 13(1) & 10(1) \\ \hline
Clua-Provost et al. [\citen{Provost2024}] & 1.0(4) & 0.091 from [\citen{Reimers2020}] & 8.2(6) $\times 10^2$ & 1.8(2) $\times 10^3$  & 41(8) & 7(2) \\ \hline
Lee et al. [\citen{lee2024intrinsichighfidelityspinpolarization}] & Not stated & 8.49 $\times 10^2$ & $< 51.44$ & 1.286 $\times 10^3$ & 22.7 & 4.54 \\ \hline
\end{tabular}
\caption{Boron vacancy electronic transition rates from the 7-level model fit to time-resolved PL data from a single sub-micron hBN flake (details of fit presented in the SI) and from previous papers.}
\label{tab:7-level_transition_rates}
\end{table}

To achieve a better fit including PL quenching behavior at high laser powers, we utilize a 9-level model that introduces coupling to an additional 2-level system (Fig. \ref{Fig_4}b). The 9-level model incorporates optical pumping ($k_{p2}$), radiative decay ($k_{r2}$), and non-radiative decay ($k_{nr}$) between the two additional levels, as well as transition rates between the 7-level and 2-level manifolds through dark conversion ($k_{di}$), dark recombination ($k_{dr}$), photoconversion ($k_{i1}$), and photorecombination ($k_{i2}$). Similar to $k_{p_0}$, $k_{p2_0}$, $k_{i1_0}$, and $k_{i2_0}$ are fit parameters that determine the scaling of the $k_{p2}$, $k_{i1}$, and $k_{i2}$ rates with laser power, respectively. The exact laser power dependencies are defined in the SI. As with the 7-level model, we fix $\gamma_0$ to a literature average of directly measured $m_s=0$ lifetimes and leave the ratio $\gamma_0/\gamma_1$ free \cite{Provost2024, Baber2021, lee2024intrinsichighfidelityspinpolarization}. We base this modified 9-level energy structure and additional transition rates on prior work that models NV charge-state photoexcitation dynamics \cite{AudeCraik, Dumeige2019}, wherein the NV$^0$ charge-state is modeled as a 2-level system connected to a 7-level NV$^-$ charge-state manifold.

Fits of time-resolved PL data to the 9-level model represent observed behavior better than fits to the 7-level model (Fig. \ref{fig:Fig_5_Rev}a vs.Fig. \ref{fig:Fig_5_Rev}b). In particular, the 9-level model replicates the observed PL quenching for the highest laser power. Quantitatively, the sum of squared residuals (SSRs) for the total dataset is lower for the 9-level model (2.13) compared to the 7-level model (3.44), demonstrating a $\sim$ 60$\%$ better fit to the data. The improved fit quality is more drastic for the highest laser power of 21.3 mW, where SSRs are more than twice as small for the 9-level model compared to the 7-level model. Fig. \ref{fig:Fig_5_Rev}c and Fig. \ref{fig:Fig_5_Rev}d show a histogram of fit residuals between the data and each model for all 5 laser powers and for the highest laser power only, respectively. In both cases, fit residuals are more sharply peaked around 0 for the 9-level model, indicating an improved fit quality, especially for the highest laser power.

Table 3 summarizes transition rates determined from 9-level model fits to PL measurements for both polarized and thermal $V_B^-$ spin distributions at several laser powers, again setting $k_{r}$ to the DFT simulation value [\citen{Reimers2020}]. Fit parameters $k_{dr}$ and $k_{i2_0}$ are found to be very small and do not materially influence the fit when left as free parameters, and thus are set to zero. We note that a non-zero recombination pathway, represented by fit parameters $k_{dr}$ and $k_{{i2}_0}$, from the 2 level system to the 7 level system is required to sustain a finite $V_B^-$ population in the steady-state; however, the sensitivity of our model and data is insufficient to place meaningful bounds on these recombination rates. Fit parameter uncertainties are calculated following the same procedure as for the 7-level model, with an additional model-dependent contribution defined as the difference in fitted parameter values between the quadratic-$k_{i1}$ and linear-$k_{i1}$ models, with the former providing a modestly better fit, as described below. Details of the fitting and uncertainty analysis are given in the SI.

\begin{table}[H]
\centering
\renewcommand{\arraystretch}{1.25}
\begin{tabular}{|c|c|}
\hline
\textbf{Electronic Transition Rate} & \textbf{9-Level Model Fit} \\ \hline
\textbf{$k_{p0}$ (MHz/mW)}    & 3.5(2) \\ \hline
\textbf{$k_{r}$ (MHz)}        & 0.091 (DFT) from [\citen{Reimers2020}] \\ \hline
\textbf{$\gamma_0$ (MHz)}     & $1.003$
 $\times 10^3$ (fixed) \\ \hline
\textbf{$\gamma_1$ (MHz)}     & $2.19(3) \times 10^3$  \\ \hline
\textbf{$\kappa_0$ (MHz)}     & 56.0(8) \\ \hline
\textbf{$\kappa_1$ (MHz)}     & $<$ 2 \\ \hline
\textbf{$k_{p2_{0}}$ (MHz/mW)} & 84(11) \\ \hline
\textbf{$k_{i1_0}$ (MHz/mW$^2$)} & $1.72(3) \times 10^{-2}$ \\ \hline
\textbf{$k_{i2_0}$ (MHz/mW)}    &0 (fixed)\\ \hline
\textbf{$k_{r2}$ (MHz)}       & $4.5(3)  \times 10^{-3}$ \\ \hline
\textbf{$k_{nr}$ (MHz)}       & $1.5(2) \times 10^3$\\ \hline
\textbf{$k_{di}$ (MHz)}       & 0.15(6)\\ \hline
\textbf{$k_{dr}$ (MHz)}       & 0 (fixed)\\ \hline
\end{tabular}
\caption{Boron vacancy electronic transition rates from the 9-level model fit to time-resolved PL data from a single sub-micron hBN flake (see fits and data in Fig. \ref{fig:Fig_5_Rev}b). Details on the fitting procedure are given in the SI.}
\label{tab:9-level_transition_rates}
\end{table}

Further work is needed to understand the physical origin of the additional 2 levels in the 9-level model, e.g., whether they are related to a different boron vacancy charge state, such as $V_B^0$. We note that modestly improved fits to PL time-course data are achieved for a quadratic laser-power dependence of photoconversion ($k_{i1})$ from the original 7-level excited states to the additional 2-level manifold. Motivated by established NV photoionization dynamics \cite{AudeCraik,Aslam_2013,PhysRevB.110.134109,ung2025highresolutiontemperatureresolvedspectroscopy}, we incorporate into the rate equation fit three possible photoconversion pathways from the ground, singlet, and excited states to the two-level manifold. Only the excited state pathway results in a non-zero photoconversion fit parameter that improves fit quality. The fitted rates of the other two pathways converge to values several orders of magnitude smaller than the excited-state photoconversion rate, and do not change fit quality when set to 0. Therefore, we fix these rates to zero in the final fit. Charge transition level calculations suggest $V_{B}^{0}$ can possibly be converted to $V_{B}^{-}$ and vice versa under photo-excitation with 515 nm illumination.\cite{WESTON2018, Strand_2020} However, recent theory work has emphasized that such ab initio charge-transition thresholds can carry substantial uncertainty and should be treated as semi-quantitative indicators rather than precise predictions.\cite{galiRecentAdvancesInitio2023} Accordingly, although the additional two levels in our 9-level description are consistent with the measured PL dynamics, we regard them only as a heuristic model. Detailed understanding of the pathways, dynamics, laser-intensity dependence, and wavelength dependence of $V_B^-$ photoconversion are important open questions for future experimental study.

\subsection{PL spectral behavior and dynamics for larger hBN flakes ($>$1 $\mu$m)}
\label{sec:largehbn}

\begin{figure*}[h]
  \centering
  \includegraphics[width=0.9 \textwidth]{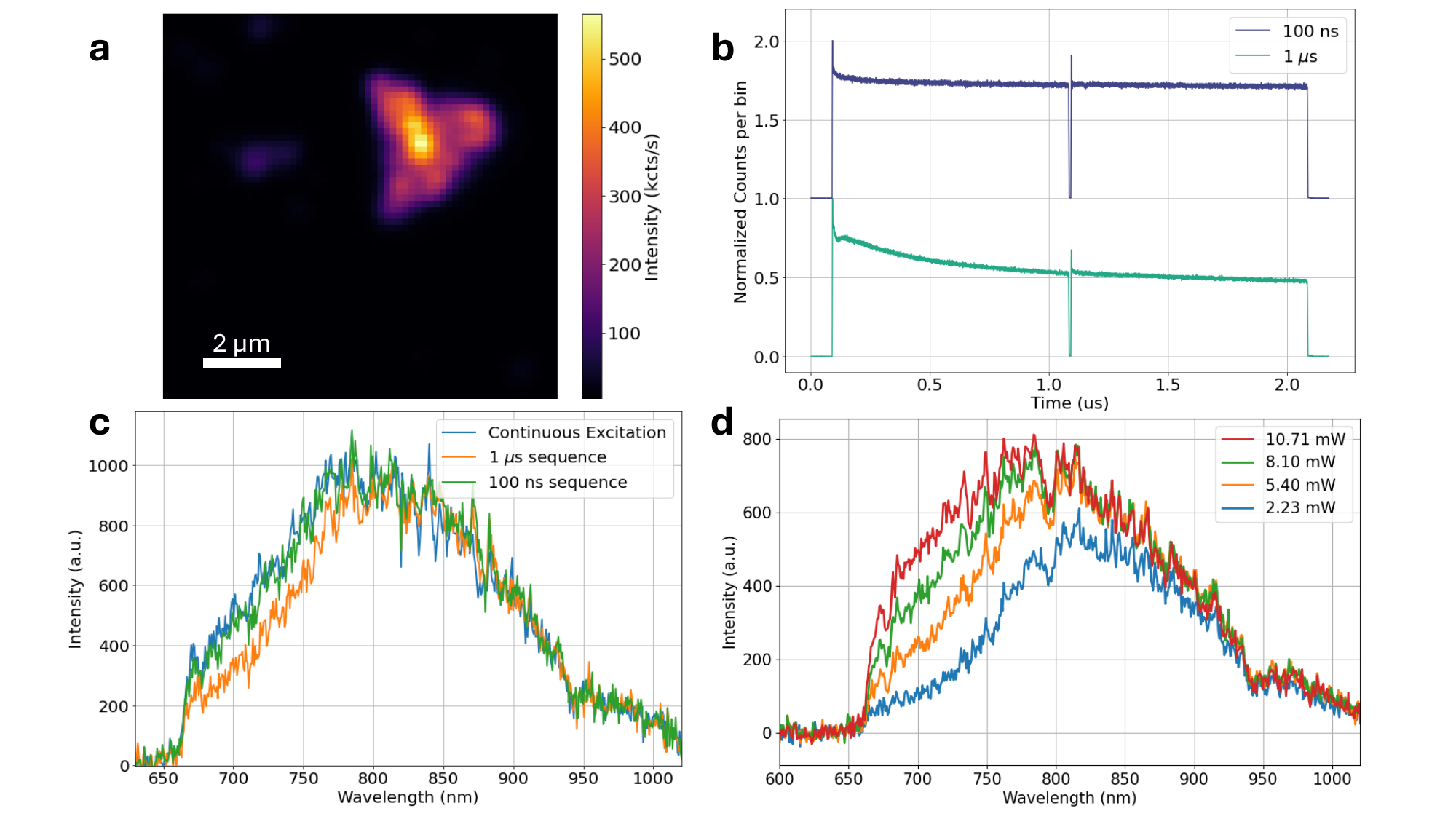}
  \caption{a) Example confocal PL map of a large ($>$1 $\mu$m) flake of neutron irradiated hBN. These larger flakes display different PL behavior from sub-micron flakes. b) Time-resolved PL measurements from a single large hBN flake for laser pulse pairs with 1 $\mu$s pulse time and dark time $t_D$ = 5 ns between pulses. The top set of signal-averaged PL measurements has a wait-time of 100 ns between pulse pairs while the lower has a 1 $\mu$s wait-time. The long decay ($\sim$1 $\mu$s) observed during the first pulse of the 1 $\mu$s wait-time sequence may result from partial conversion of $V^{-}_{B}$ to $V^{0}_{B}$ within the ensemble of defects. c) PL emission spectra for continuous laser excitation, the 1 $\mu$s pulse pair wait-time sequence, and the 100 ns pulse pair wait-time sequence, with all measurements performed using 11.6 mW laser power. Note the observed increase in PL in the $\sim$675-775 nm range for continuous excitation and the 100 ns pulse pair wait-time sequence, potentially due to enhanced conversion of $V^{-}_{B}$ to $V_B^0$. This increase is not observed for the 1 $\mu$s sequence, potentially because this longer pulse pair wait-time is sufficient for recombination back to $V^{-}_{B}$. d) PL emission spectra of a large hBN flake for 4 laser excitation powers. In both c) and d), the sharp cut-off observed at $\sim$660 nm is from the dichroic mirror (Thorlabs DMLP650R).}
  \label{Fig_5}
\end{figure*}

We observe distinctly different photophysical behavior in neutron-irradiated hBN flakes with size greater than $\sim$1 $\mu$m. Figure \ref{Fig_5}a shows a PL confocal scan of one such large flake. We characterize the PL behavior of a single large hBN flake by applying two different laser pulse-pair sequences, with 1 $\mu$s pulse time, dark time $t_D =5 $ ns between pulses, and wait-times between pulse pairs of either 100 ns or 1 $\mu$s. We perform time-resolved PL measurements using these pulse-pair sequences, while also monitoring the PL emission spectrum. The resulting PL time traces (signal averaged over $\sim$5 minutes for each wait-time sequence) are shown in Fig. \ref{Fig_5}b. For the 1 $\mu$s wait-time, we observe a long PL decay ($\sim$1 $\mu$s)  that is not observed for the 100 ns wait-time. Full PL recovery after this long decay is on the order of several hundreds of ns.

For continuous laser excitation and the 100 ns pulse pair wait-time sequence, we observe an increase in PL emission as a function of laser power in the range of $\sim$675-775 nm, when compared to measurements for the 1 $\mu$s pulse pair wait-time sequence (see Fig. \ref{Fig_5}c and \ref{Fig_5}d). Each of these measurements are performed using a broadband optical spectrometer with about 1 minute of signal averaging. This behavior contrasts with that observed in sub-micron hBN flakes, for which no significant laser-power dependence is observed in the normalized PL emission spectrum (see Fig. S3 of the SI). Our working hypothesis to explain these observations is that a laser-power dependent increase in PL $\sim$675-775 nm in larger hBN flakes arises from enhanced conversion of $V^{-}_{B}$ to $V^{0}_{B}$ (or perhaps another charge state) under continuous laser excitation and the 100 ns pulse pair wait-time sequence; whereas for the 1 $\mu$s wait-time sequence in larger hBN flakes --- and also for sub-micron flakes under any of the laser excitation protocols --- there is sufficient time for charge-state recombination back to $V^{-}_{B}$, and hence no enhanced PL $\sim$675-775 nm. This phenomenon requires further investigations to fully understand the underlying mechanisms, which could arise from strain, aggregates of smaller hBN flakes, or edge effects, e.g., measurements of PL behavior as a function of temperature and laser wavelength, as well as hBN flake size and defect/material preparation.

\section{Conclusion}
In summary, we employ a nanosecond rise-time 515 nm laser to perform a direct measurement of the singlet lifetime of an ensemble of $V^{-}_{B}$ (boron vacancy) defects in neutron-irradiated, sub-micron flakes of hBN. We repeat the measurement on 16 hBN flakes at room temperature and obtain a mean lifetime $\tau_s = 15(3)$ ns. For a single sub-micron hBN flake, we measure photoluminescence (PL) time traces for thermal and polarized electronic spin distributions of the boron vacancy ensemble. Fitting the PL time-course data to a 9-level model of  of $V_{B}^{-}$ electronic and spin levels, we extract more accurate values for the electronic energy level transition rates than for the conventional 7-level model, potentially indicating optically-induced conversion to $V^{0}_{B}$ or another state. We also observe distinct PL behavior in large ($>$1 $\mu$m) neutron-irradiated hBN flakes that may result from optically-induced changes in boron vacancy charge state. This behavior includes a long ($\sim$1 $\mu$s) PL decay for an extended wait-time ($\sim$1 $\mu$s) between laser pulse-pairs; and an increase in PL emission as a function of excitation laser power in the range $\sim$675-775 nm for continuous laser excitation or for a short (100 ns) wait-time between laser pulse-pairs. These measurements add to the growing understanding of the $V^{-}_{B}$ defect in hBN.

\section{Experimental Methods}

\subsection{Sample preparation}

hBN samples (micro-powder from 2D Semiconductors, USA) are neutron-irradiated in a Triga Mark IPR-R1 nuclear reactor (CDTN, Brazil), with thermal neutron flux of $4\times 10^{12}\,\rm{n}\, \rm{cm}^{ - 2} \,\rm{s}^{-1}$ up to a total dose of $2.3 \times 10^{17} \, \rm{n} \, \rm{cm}^{ - 2}$ (16 h). Considering that thermal neutrons are blocked by a Cd capsule, and the fast and epithermal neutron flux is about one order of magnitude lower, the total neutron dose to the hBN samples is about $2.3 \times 10^{16} \, \rm{n} \, \rm{cm}^{ - 2}$. Compared to electron, proton, and ion irradiations, fast neutrons have no electrical charge, and therefore present homogeneous and high penetration depths in almost all materials.\cite{HOSSEINI2021} Irradiated hBN samples are pre-characterized by electron paramagnetic resonance (EPR) in the 10 to 300 K temperature range, from which the presence of $V^{-}_{B}$ defects is inferred.\cite{Toledo2020} Irradiated samples, consisting of flakes up to 5 $\mu$m in size, are sonicated in solution and dropcast on a glass cover slip.

\subsection{Photoluminescence (PL) measurement setup}

Measurements are performed using a home-built confocal microscope with a digitally modulated fast rise-time 515 nm laser (H{\"u}bner Cobolt 06-MLD). To control the laser modulation, we use a Pulse Streamer 8/2 (Swabian Instruments) along with a TTL buffer (Texas Instrument SN64BCT25244DW) to raise the pulse signal from 2.6 V to 5 V, which is the optimal voltage to drive the Cobolt 06-MLD. To excite boron vacancy photoluminescence (PL) from an hBN sample, a dichroic mirror (Thorlabs DMLP650R) is used to steer the incident laser beam to an infrared objective (0.7 NA OptoSigma PAL-100-NIR-HR). 
PL is focused through a pinhole and split with a 50-50 beam splitter (Thorlabs CCM1-BS014). 
All mirrors after the pinhole are broadband dielectric mirrors (Thorlabs BB1-E03-10). 
One path is filtered using a 700 nm long pass (Thorlabs FEL0700) and a 900 nm short pass (Thorlabs FESH0900) and coupled via a fiber to a photodetector (Excelitas SPCM-AQRH-W5-FC). 
The other path is filtered with a 550 nm long pass and directed to a spectrometer (Ocean Optics QE Pro). Time-resolved PL measurements are performed using a Swabian Time Tagger 20 to record pulses from the photodetector. 

\medskip
\textbf{Supporting Information} \par 
Supporting Information (SI) is available online.

\medskip
\textbf{Acknowledgements} \par 

We thank Alessandro Restelli for technical assistance regarding electronics in the experiment; and Sean Blakley and Kristine Ung for helpful discussions. K. Krambrock thanks the Brazilian agencies CNPq and Fapemig for financial support. This work is supported by, or in part by, the U.S. Air Force Office of Scientific Research under Grant No. FA9550-22-1-0312 as part of the MURI project Comprehensive Minimally/non-invasive Multifaceted Assessment of Nano-/microelectronic Devices (CoMMAND); the National Science Foundation Graduate Research Fellowship Program under Grant No. DGE 2236417; the Laboratory for Physical Sciences Jumping Electron Quantum Fellowship Program under Award No. H9823022C0029; the U.S. Army Research Laboratory under Contract No. W911NF2420143; the Maryland Procurement Office under Award No. H9823019C0220; and the University of Maryland Quantum Technology Center.

\medskip
\bibliographystyle{MSP}
\bibliography{references}

\end{document}